\documentclass[aps,twocolumn,showpacs,prl,superscriptaddress]{revtex4}
\usepackage{amssymb}
\usepackage{natbib}
\usepackage{amsmath}
\usepackage{amsfonts}
\usepackage{graphicx}
\usepackage{mathrsfs}
\usepackage{dcolumn}
\usepackage{bm}
\usepackage{color}
\usepackage{cancel}
\usepackage{mathbbol}
\usepackage{epsfig}
\usepackage{units}
\usepackage{hyperref}

\definecolor{rot}{rgb}{0.75,0.05,0.25}
\definecolor{hellgrau}{gray}{0.5}
\definecolor{blau}{rgb}{0,0,0.7}

\begin{document}

\title{Reply to M. Mel\'endez  and W. G. Hoover [arXiv:1206.0188v2]}
\author{Michele Campisi}
\affiliation{Institute of Physics, University of Augsburg,
  Universit\"atsstr. 1, D-86135 Augsburg, Germany}
\author{Fei Zhan}
\affiliation{International Center for Quantum Materials, Peking University, 100871, Beijing,
China}
\author{Peter Talkner}
\affiliation{Institute of Physics, University of Augsburg,
  Universit\"atsstr. 1, D-86135 Augsburg, Germany}
\author{Peter H\"anggi}
\affiliation{Institute of Physics, University of Augsburg,
  Universit\"atsstr. 1, D-86135 Augsburg, Germany}
\date{\today }

\begin{abstract}
In response to the recent critical comment by M. Mel\'endez  and W. G. Hoover [arXiv:1206.0188v2] on our work
[M. Campisi \emph{et al.}, Phys. Rev. Lett. \textbf{108}, 250601 (2012)], we show that their molecular dynamics simulations do not disprove our theory but in fact convincingly corroborate it.
\end{abstract}


 \maketitle

In their comment \cite{MelendezArXiv1206.0188} to our Letter \cite{Campisi12PRL108},
Mel\'endez  and Hoover
claim that the Hamiltonian thermostat presented in \cite{Campisi12PRL108},
has ``very unusual drawbacks'' that ``make it impractical for many applications''.
In support of this statement they present
molecular dynamics simulations.
We show here that, quite on the contrary, those simulations corroborate our theory.

For convenience here we reproduce Figure 1 of the comment of Mel\'endez  and Hoover
\cite{MelendezArXiv1206.0188}, see Fig. 1. The figure shows the numerically computed energy probability distribution of one particle (blue symbols) and two particles (black symbols) in a 1D box interacting with a log-oscillator.
\begin{figure}
\begin{center}
\includegraphics[width=.48\textwidth]{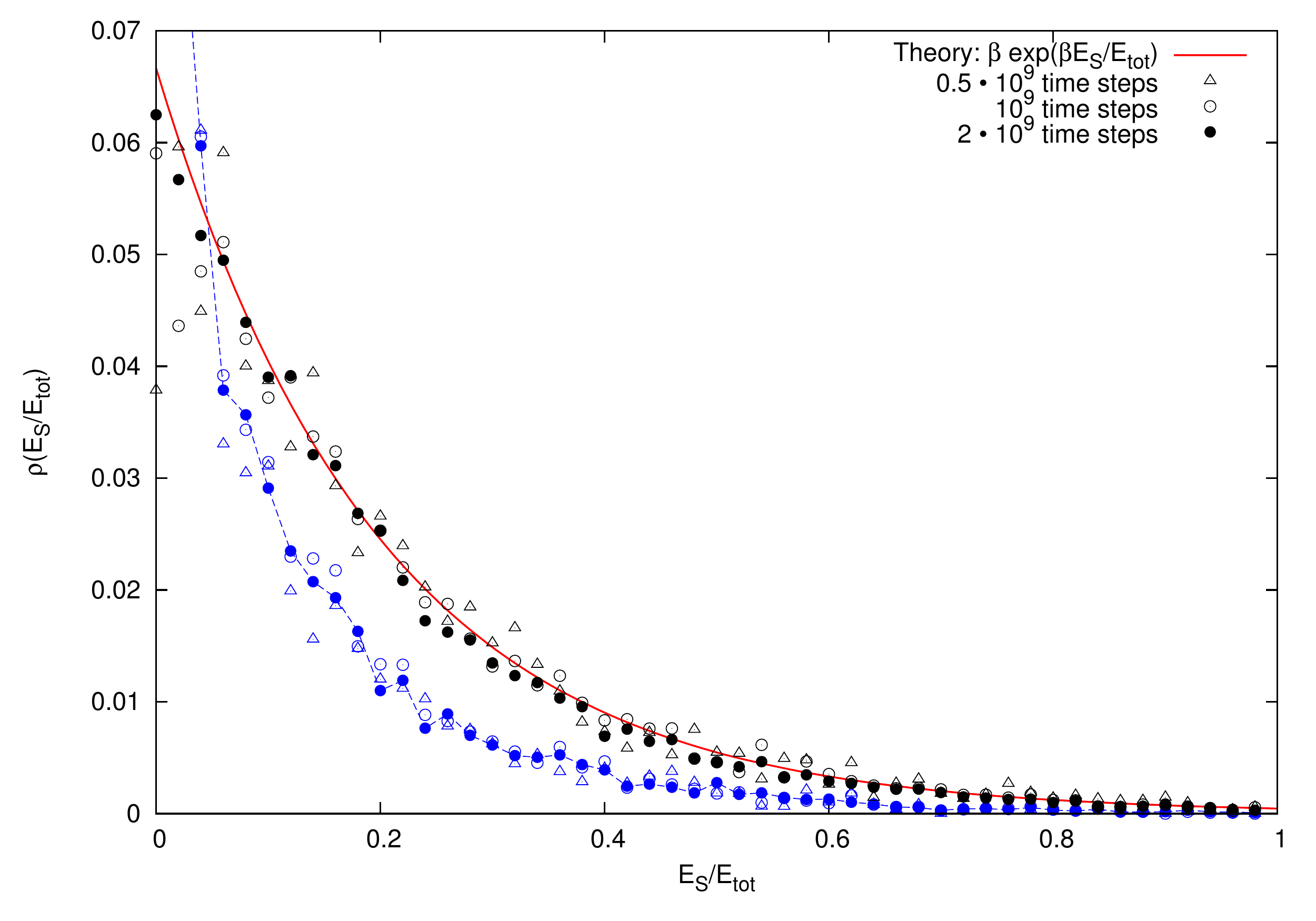}
\caption{Figure 1 of the comment of Mel\'endez  and Hoover, Ref. \cite{MelendezArXiv1206.0188}}
\end{center}
\end{figure}
In the caption the authors state:
\begin{quote}
``The blue points correspond to a system of only one thermostated particle, where the results failed to converge to the theoretical prediction during the simulation run.''
\end{quote}
Mel\'endez  and Hoover make a mistake in calculating the ``theoretical prediction''
in the case of one particle.
The theory predicts that the probability density function is the Gibbs distribution:
\begin{eqnarray}
 \rho(\mathbf{q},\mathbf{p})={e^{-H_S(\mathbf{q},\mathbf{p})/T}
}/{Z(T)} \, ,
\label{eq:canonical}
\end{eqnarray}
where $Z(T)$ is the partition function, see Eq. (7) of our Letter. In the energy space the Gibbs distribution function reads, as shown in statistical mechanics textbooks \cite{HuangBook}, and also mentioned in
our Letter \cite{Campisi12PRL108},
\begin{eqnarray}
 \rho(E_S)=\frac{e^{-E_S/T}\Omega_S(E_S)
}{Z(T)} \, ,
\end{eqnarray}
where $\Omega_S(E_S)$ is the density of states of the system.
\begin{figure}[b]
\begin{center}
\includegraphics[width=.48\textwidth]{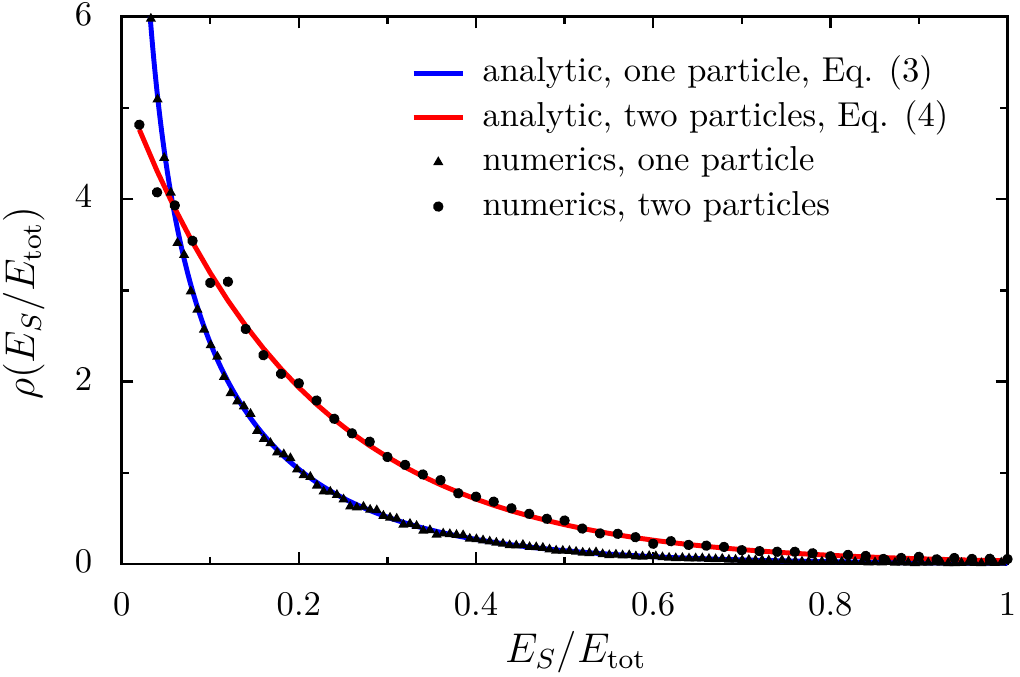}
\caption{Our own simulations with one and two thermostated particles in a 1D box, compared with the correct theoretical predictions, Eq. (3) and (4), respectively.}
\end{center}
\end{figure}
It is well known that the density of states of a system with a Hamiltonian consisting in the sum
of $n$ quadratic terms is of the form $\Omega_S(E_S) \propto E_S^{n/2-1}$ \cite{HuangBook}.
For one particle in a 1D box, $n=1$, and $\Omega_S(E_S)\propto E_S^{-1/2}$.
For two particles in a 1D box, $n=2$, and $\Omega_S(E_S)$ is a constant, as we explicitly said in our Letter
\cite{Campisi12PRL108}. 
Therefore the (blue) data in Fig. 1 from a simulation with {one} particle should be compared with
\begin{eqnarray}
 \rho(E_S)=\frac{e^{-E_S/T}E_S^{-1/2}
}{\int_0^{\infty} e^{-E_S/T} E_S^{-1/2}\mathrm{d}E_S} \qquad \text{(one particle)}
\label{eq:canonical1p}
\end{eqnarray}
and the (black) data  in Fig. 1 from a simulation with {two} particles should be compared with
\begin{eqnarray}
 \rho(E_S)=\frac{e^{-E_S/T}
}{\int_0^{\infty} e^{-E_S/T} \mathrm{d}E_S} \qquad \text{(two particles)}
\label{eq:canonical2p}
\end{eqnarray}
that is the red line in Fig. 1.

Mel\'endez  and Hoover mistakenly compare the one-particle data sets with Eq. (4), which pertains
instead to the case of two particles only.

Rather than evidencing a ``very unusual drawback" of our work, the simulations of Mel\'endez  and Hoover corroborate our theory. See our own simulations in Fig. 2 compared with the
correct theoretical predictions, Eqs. (3) and (4) respectively. In comparing our simulations in Fig. 2, with Hoover and Melendez simulations in Fig. 1, note that the main difference is the scale of the vertical axis. This is because, unlike 
Mel\'endez  and Hoover, we have properly normalized the data so that the area below the curves is $1$, as we did in our Letter \cite{Campisi12PRL108}.

In Fig. 2 of their comment \cite{MelendezArXiv1206.0188}, Mel\'endez  and Hoover provide
the results from a simulation of a 1D chain of eighteen quartic oscillators, which are linearly coupled  
to two log-oscillators of same strength $T$. 
Actually that figure demonstrates a convergence of the particle temperatures toward the value given by the log-oscillators strength $T$, apparently in agreement with our theory. 
In Fig. 3  of their comment \cite{MelendezArXiv1206.0188}, Mel\'endez  and Hoover provide
the results from a similar simulation but for a nonequilibrium scenario with the two log-oscillators having different strengths $T_1$ and $T_2$.
These simulations with linear chains, are neither sufficiently documented,
nor conclusive. It is not possible to infer whether the simulations
were done in a proper parameter regime and to draw any conclusions from them.   
The question whether and under which conditions log-oscillators may be employed to simulate non-equilibrium situations
is off-topic with respect to the focus of our Letter  \cite{Campisi12PRL108}, and needs further thorough investigations.

To sum up, the claim of Mel\'endez  and Hoover that logarithmic oscillators
``are not very useful in most practical applications, whether simulations or experiments'' has no scientific foundation.


\end{document}